\documentclass[12pt,letterpaper,english,onecolumn]{IEEEtran}
\usepackage[T1]{fontenc}
\usepackage[latin9]{inputenc}
\usepackage{color}
\usepackage{babel}
\usepackage{float}
\usepackage{amsmath}
\usepackage{amsthm}
\usepackage{amssymb}
\usepackage{graphicx}
\usepackage{setspace}
\usepackage[unicode=true,pdfusetitle,
 bookmarks=true,bookmarksnumbered=false,bookmarksopen=false,
 breaklinks=false,pdfborder={0 0 1},backref=false,colorlinks=false]
 {hyperref}

\makeatletter

\pdfpageheight\paperheight
\pdfpagewidth\paperwidth

\providecommand{\tabularnewline}{\\}
\floatstyle{ruled}
\newfloat{algorithm}{tbp}{loa}
\providecommand{\algorithmname}{Algorithm}
\floatname{algorithm}{\protect\algorithmname}

\theoremstyle{plain}
\newtheorem{thm}{\protect\theoremname}
\theoremstyle{remark}
\newtheorem{rem}[thm]{\protect\remarkname}
\theoremstyle{plain}
\newtheorem{cor}[thm]{\protect\corollaryname}

 \usepackage{cite}

\makeatother

\providecommand{\corollaryname}{Corollary}
\providecommand{\remarkname}{Remark}
\providecommand{\theoremname}{Theorem}

\begin{document}

\title{Dual Link Algorithm for the Weighted Sum Rate Maximization in MIMO
Interference Channels}

\author{Xing Li\textsuperscript{1}, Seungil You\textsuperscript{2}, Lijun
Chen\textsuperscript{3}, An Liu\textsuperscript{4}, Youjian (Eugene)
Liu\textsuperscript{1\thanks{This work was supported in part by NSF grants ECCS-1408604, IIP-1414250.}}\\
{\small{}\textsuperscript{1}Department of Electrical, Computer, and
Energy Engineering, University of Colorado at Boulder}\\
{\small{}\textsuperscript{2}Department of Computing and Mathematical
Sciences, California Institute of Technology}\\
{\small{}\textsuperscript{3}Department of Computer Science, University
of Colorado at Boulder}\\
{\small{}\textsuperscript{4}Department of Electronic and Computer
Engineering, Hong Kong University of Science and Technology}}
\maketitle
\begin{abstract}
MIMO interference network optimization is important for increasingly
crowded wireless communication networks. We provide a new algorithm,
named Dual Link algorithm, for the classic problem of weighted sum-rate
maximization for MIMO multiaccess channels (MAC), broadcast channels
(BC), and general MIMO interference channels with Gaussian input and
a total power constraint. For MIMO MAC/BC, the algorithm finds optimal
signals to achieve the capacity region boundary. For interference
channels with Gaussian input assumption, two of the previous state-of-the-art
algorithms are the WMMSE algorithm and the polite water-filling (PWF)
algorithm. The WMMSE algorithm is provably convergent, while the PWF
algorithm takes the advantage of the optimal transmit signal structure
and converges the fastest in most situations but is not guaranteed
to converge in all situations. It is highly desirable to design an
algorithm that has the advantages of both algorithms. The dual link
algorithm is such an algorithm. Its fast and guaranteed convergence
is important to distributed implementation and time varying channels.
In addition, the technique and a scaling invariance property used
in the convergence proof may find applications in other non-convex
problems in communication networks.\end{abstract}


\section{Introduction\label{sec:Introduction}}

One of the main approaches to accommodating the explosive growth in
mobile data is to reduce the cell size and increase the base station
or access point density, while all cells reuse the same frequency
spectrum. However, the inter-cell interference becomes severe because
the probability of line of sight increases as cell size shrinks. On
the other hand, the situation is not hopeless. As promised by interference
alignment through joint transmit signal design, every user can have
half of the bandwidth at infinite SNR \cite{Jafar_2008ITIT_InterferenceAlignmentDegreesFreedomUserInterferenceChannel}.
Consequently, joint transmit signal design algorithms are expected
to be employed to manage interference, or equivalently, maximize data
rate at practical SNR, and asymptotically achieve interference alignment.
The main hurdle to joint transmit signal design is the collection
of global channel state information (CSI) and coordination/feedback
overhead. 

In this paper, we design a new algorithm, named Dual Link algorithm,
that jointly optimizes the covariance matrices of transmit signals
of multiple transmitters in order to maximize the weighted sum-rate
of the data links. The algorithm is ideally suited for distributed
implementation where only local channel state information is needed.
The algorithm works for the MIMO B-MAC networks and assumes Gaussian
transmit signal. The MIMO B-MAC network model \cite{Liu_IT10s_Duality_BMAC}
includes broadcast channel (BC), multiaccess channel (MAC), interference
channels, X networks, and many practical wireless networks as special
cases. The weighted sum-rate maximization can be used for other utility
optimization by finding appropriate weights and thus is a classic
problem to solve. The problem is non-convex, and various algorithms
have been proposed for various cases, e.g., \cite{Caire_09ISIT_BC_linear_constraints,Caire_09sTSP_BC_intercell_interference,Jindal_IT05_IFBC,Giannakis_TIT11_IFPMIMO,Liu_IT10s_Duality_BMAC,Berry_MonoIFCpricing_ISIT09,Luo_TSP11_WMMSE,Yu_IT04_MIMO_MAC_waterfilling_alg,Weiyu_IT06_DualIWF,Wei_07ITW_MultiuserWF,Zhang_09ISIT_BC_MAC_duality_linear_constraints,ZhangLan_09TWC_Weighted_rate_BC}.
Among the previous state-of-the-art algorithms, we \textcolor{black}{have
proposed the }polite water-filling\textcolor{black}{{} (PWF) algorithm}
\cite{Liu_IT10s_Duality_BMAC}\textcolor{black}{. }Because it takes
advantage of the optimal transmit signal structure for an achievable
rate region, the polite water-filling structure, the PWF algorithm
has the lowest complexity and the fastest convergence when it converges.
However, in some strong interference cases, it has small oscillation.
Another excellent algorithm is the WMMSE algorithm in \cite{Luo_TSP11_WMMSE}.
It was proposed for beamforming matrix design for the MIMO interfering
broadcast channels but could be readily applied to the more general
B-MAC networks and input covariance matrix design. It transforms the
weighted sum-rate maximization into an equivalent weighted sum mean
square error minimization problem, which has three sets of variables
and is convex when any two variable sets are fixed. With the block
coordinate optimization technique, the WMMSE algorithm is guaranteed
to converge to a stationary point, though the convergence is observed
in simulations to be slower than the PWF algorithm. 

It is thus highly desirable to have an algorithm with the advantages
of both PWF and WMMSE algorithms, i.e., fast convergence by taking
advantage of the optimal transmit signal structure and provable convergence
for the general interference network. The main contribution of this
paper is such an algorithm, the dual link algorithm. It exploits the
forward-reverse link rate duality in a new way. Numerical experiments
demonstrate that the dual link algorithm is almost as fast as the
PWF algorithm and can be a few iterations or more than ten iterations
faster than the WMMSE algorithm, depending on the desired accuracy
with respect to the local optimum. Note that being faster even by
a couple iterations will be critical in distributed implementation
in time division duplex (TDD) networks with time varying channels,
where the overhead of each iteration costs significant signaling resources
between the transmitters and the receivers. The faster the convergence
is, the faster channel variations can be accommodated by the algorithm.
Indeed, the dual link algorithm is highly scalable and suitable for
distributed implementation because, for each data link, only its own
channel state and the aggregated interference plus noise covariance
need to be estimated no matter how many interferers are there. We
also show that the dual link algorithm can be easily modified to deal
with systems with colored noise.

Another contribution of this paper is the proof of the monotonic convergence
of the algorithm. It uses only very general convex analysis, as well
as a particular scaling invariance property that we identify for the
weighted sum-rate maximization problem. We expect that the scaling
invariance holds for and our proof technique applies to many non-convex
problems in communication networks. 

The centralized version of dual link algorithm for total power constraint
has been generalized to multiple linear constraints using a minimax
approach \cite{ChenL_CoRR2013_WSR_MIMO_minimax}, and has stimulated
the design of another monotonic convergent algorithm based on convex-concave
procedure \cite{YouS_14sGlobecom_cvxccv_short} which has slower convergence
but can handle nonlinear convex constraints. Nevertheless, the dual
link algorithm uses a different derivation approach, which is based
on the optimal transmit signal structure, and easily leads to a low
complexity distributed algorithm. Thus, the special case of total
power constraint provides a different view and insight than the general
multiple linear constraint case.

The rest of this paper is organized as follows. Section \ref{sec:Preliminaries}
presents the system model, formulates the problem, and briefly reviews
the related results on the rate duality and polite water-filling structure.
Section \ref{sec:DL_alg} proposes the new algorithm and establishes
its monotonic convergence. Section \ref{sec:Colored-Noise} shows
how to modify the dual link algorithm for the environment with colored
noise and discusses distributed implementation. Numerical examples
are presented in Section \ref{sec:Simulation-Results}. Complexity
analysis is provided in Section \ref{sec:Complexity-Analysis}. Section
\ref{sec:Conclusion} concludes.

\section{Preliminaries\label{sec:Preliminaries}}

In this section, we describe the system model and formulate the optimization
problem, then briefly review some related results on the polite water-filling,
which leads to the design of the dual link algorithm.

\begin{figure}
\begin{centering}
\includegraphics[width=100mm]{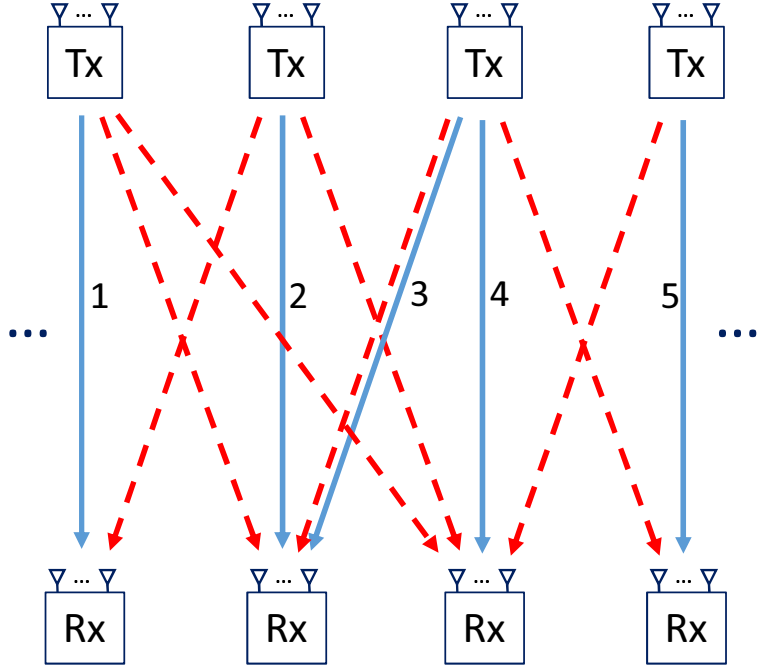}
\par\end{centering}

\centering{}\caption{\label{fig:BMAC_ex}An example of B-MAC network. The solid lines represent
data links and the dash lines represent interference.}
\end{figure}

\subsection{B-MAC Interference Networks\label{sub:B-MAC-Interference-Networks}}

We consider a general interference network named\textcolor{black}{{}
MIMO B-MAC network} \textcolor{black}{with multiple transmitters and
receivers\cite{Liu_IT10s_Duality_BMAC,Liu_10sTSP_Fairness_rate_polit_WF}.
A transmitter in the MIMO B-MAC network may send independent data
to different receivers, like in BC, and a receiver may receive independent
data from different transmitters, like in MAC. }Assume there are totally
$L$ mutually interfering data links in a B-MAC network. Link $l$'s
physical transmitter is $T_{l}$, which has $L_{T_{l}}$ many antennas.
Its physical receiver is $R_{l}$, which has $L_{R_{l}}$ many antennas.
Figure \ref{fig:BMAC_ex} shows an example of B-MAC networks with
five data links. Link 2 and 3 have the same physical receiver. Link
3 and 4 have the same physical transmitter. When multiple data links
have the same receiver or the same transmitter, interference cancellation
techniques such as successive decoding and cancellation or dirty paper
coding can be applied \cite{Liu_IT10s_Duality_BMAC}. The received
signal at $R_{l}$ is
\begin{eqnarray}
\mathbf{y}_{l} & = & \sum_{k=1}^{L}\mathbf{H}_{l,k}\mathbf{x}_{k}+\mathbf{n}_{l},\label{eq:recvsignal}
\end{eqnarray}
where $\mathbf{x}_{k}\in\mathbb{C}^{L_{T_{k}}\times1}$ is the transmit
signal of link $k$ and is modeled as a circularly symmetric complex
Gaussian vector; $\mathbf{H}_{l,k}\in\mathbb{C}^{L_{R_{l}}\times L_{T_{k}}}$
is the channel state information (CSI) matrix between $T_{k}$ and
$R_{l}$; and $\mathbf{n}_{l}\in\mathbb{C}^{L_{R_{l}}\times1}$ is
a circularly symmetric complex Gaussian noise vector with identity
covariance matrix. The circularly symmetric assumption of the transmit
signal can be dropped easily by applying the proposed algorithm to
real Gaussian signals with twice the dimension. Multiple channel uses
can be combined into a larger B-MAC networks with parallel channels,
like in interference alignment \cite{Jafar_2008ITIT_InterferenceAlignmentDegreesFreedomUserInterferenceChannel}.

\subsection{Problem Formulation}

Assuming the channels are known at both the transmitters and receivers
(CSITR), an achievable rate of link $l$ is
\begin{equation}
\mathcal{I}_{l}\left(\mathbf{\Sigma}_{1:L}\right)=\textrm{log}\left|\mathbf{I}+\mathbf{H}_{l,l}\mathbf{\Sigma}_{l}\mathbf{H}_{l,l}^{\dagger}\mathbf{\Omega}_{l}^{-1}\right|\label{eq:achrate}
\end{equation}
where $\mathbf{\Sigma}_{l}$ is the covariance matrix of $\mathbf{x}_{l}$;
and $\mathbf{\Omega}_{l}$ is the interference-plus-noise covariance
matrix of the $l^{\text{th}}$ link,
\begin{equation}
\mathbf{\Omega}_{l}=\mathbf{I}+\sum_{k=1,k\ne l}^{L}\mathbf{H}_{l,k}\mathbf{\Sigma}_{k}\mathbf{H}_{l,k}^{\dagger}.\label{eq:Omega}
\end{equation}
 If the interference from link $k$ to link $l$ is completely canceled
using successive decoding and cancellation or dirty paper coding,
we can simply set $\mathbf{H}_{l,k}=\mathbf{0}$ in (\ref{eq:Omega}).
Otherwise, the interference is treated as noise. This allows this
paper to cover a wide range of communication techniques.

The optimization problem that we want to solve is the weighted sum-rate
maximization under a total power constraint:
\begin{eqnarray}
\text{\textbf{WSRM\_TP}:} & \max_{\mathbf{\Sigma}_{1:L}} & \sum_{l=1}^{L}w_{l}\mathcal{I}_{l}\left(\mathbf{\Sigma}_{1:L}\right)\label{eq:WSRM_TP}\\
 & \text{s.t.} & \mathbf{\Sigma}_{l}\succeq0\text{, }\forall l,\nonumber \\
 &  & \sum_{l=1}^{L}\text{Tr}\left(\mathbf{\Sigma}_{l}\right)\le P_{\text{T}},\nonumber 
\end{eqnarray}
where $w_{l}>0$ is the weight for link $l$. The generalization to
multiple linear constraints as in \cite{Liu_10sTSP_MLC} is given
in \cite{ChenL_CoRR2013_WSR_MIMO_minimax}, which covers the individual
power constraints as a special case.

\subsection{Rate Duality and Polite Water-filling}

We review the relevant results on the non-convex optimization (\ref{eq:WSRM_TP})
given in \cite{Liu_IT10s_Duality_BMAC}. Dual network, reverse links,
and rate duality were introduced. The optimal structure of the transmit
signal covariance matrices is polite water-filling structure, whose
definition involves the reverse link interference plus noise covariance
matrices. It suggests an iterative polite water-filling algorithm,
which is compared with the new algorithm in this paper. The polite
water-filling structure was used to derive a dual transformation,
based on which the new algorithm in this paper has been designed.

\subsubsection*{A Dual Network and the Reverse Links}

A virtual dual network can be created from the original B-MAC network
by reversing the roles of all transmitters and receivers and replacing
the channel matrices with their conjugate transpose. The data links
in the original networks are denoted as \emph{forward links} while
those in the dual network are denoted as \emph{reverse links}. We
use\textcolor{black}{{} $\hat{}$ to denote the corresponding terms
in the }reverse links. The interference-plus-noise covariance matrix
of reverse link $l$ is
\begin{eqnarray}
\hat{\mathbf{\Omega}}_{l} & = & \mathbf{I}+\sum_{k=1,k\ne l}^{L}\mathbf{H}_{k,l}^{\dagger}\hat{\mathbf{\Sigma}}_{k}\mathbf{H}_{k,l},\label{eq:Omega_hat}
\end{eqnarray}
 where $\hat{\mathbf{\Sigma}}_{k}$ is the transmit signal covariance
matrix of reverse link $k$. The achievable rate of reverse link $l$
is
\begin{equation}
\mathcal{\hat{I}}_{l}\left(\hat{\mathbf{\Sigma}}_{1:L}\right)=\textrm{log}\left|\mathbf{I}+\mathbf{H}_{l,l}^{\dagger}\hat{\mathbf{\Sigma}}_{l}\mathbf{H}_{l,l}\hat{\mathbf{\Omega}}_{l}^{-1}\right|.\label{eq:rate_reverse}
\end{equation}
 A dual optimization problem corresponding to \ref{eq:WSRM_TP} can
formulated as
\begin{eqnarray}
\text{\textbf{WSRM\_TP\_D}:} & \max_{\mathbf{\Sigma}_{1:L}} & \sum_{l=1}^{L}w_{l}\hat{\mathcal{I}}_{l}\left(\hat{\mathbf{\Sigma}}_{1:L}\right)\label{eq:WSRM_DTP}\\
 & \text{s.t.} & \hat{\mathbf{\Sigma}}_{l}\succeq0\text{, }\forall l,\nonumber \\
 &  & \sum_{l=1}^{L}\text{Tr}\left(\hat{\mathbf{\Sigma}}_{l}\right)\le P_{\text{T}}.\nonumber 
\end{eqnarray}

\subsubsection*{\textcolor{black}{Rate Duality}}

\textcolor{black}{The rate duality states that the achievable rate
regions of the forward link channels $\left(\left[\mathbf{H}_{l,k}\right]\text{, }\sum_{l=1}^{L}\text{Tr}\left(\mathbf{\Sigma}_{l}\right)\le P_{\text{T}}\right)$
and reverse link channels $\left(\left[\mathbf{H}_{k,l}^{\dagger}\right]\text{, }\sum_{l=1}^{L}\text{Tr}\left(\hat{\mathbf{\Sigma}}_{l}\right)\le P_{\text{T}}\right)$
are the same \cite{Liu_IT10s_Duality_BMAC}. The achievable rate regions
are defined using rates in (\ref{eq:achrate}) and (\ref{eq:rate_reverse}).
A }\textit{covariance transformation in}\textcolor{black}{{} \cite{Liu_IT10s_Duality_BMAC}
calculates the reverse link input covariance matrices $\hat{\mathbf{\Sigma}}_{l}$'s
from the forward ones $\mathbf{\Sigma}_{l}$'s. }The rate duality
is proved by showing that these calculated \textcolor{black}{$\hat{\mathbf{\Sigma}}_{l}$'s}
achieves equal or higher rates than the forward link rates employing
\textcolor{black}{$\mathbf{\Sigma}_{l}$'s} under the same value of
power constraint $P_{\text{T}}$\textcolor{black}{{} \cite{Liu_IT10s_Duality_BMAC}.}

\subsubsection*{Polite Water-filling Structure}

We review the polite water-filling results from \cite{Liu_IT10s_Duality_BMAC}.
The Lagrange function of problem (\ref{eq:WSRM_TP}) is
\begin{eqnarray*}
 &  & L\left(\mu,\mathbf{\Theta}_{1:L},\mathbf{\Sigma}_{1:L}\right)\\
 & = & \sum_{l=1}^{L}w_{l}\textrm{log}\left|\mathbf{I}+\mathbf{H}_{l,l}\mathbf{\Sigma}_{l}\mathbf{H}_{l,l}^{\dagger}\mathbf{\Omega}_{l}^{-1}\right|+\sum_{l=1}^{L}\textrm{Tr}\left(\mathbf{\Sigma}_{l}\mathbf{\Theta}_{l}\right)\\
 &  & +\mu\left(P_{T}-\sum_{l=1}^{L}\text{Tr}\left(\mathbf{\Sigma}_{l}\right)\right),
\end{eqnarray*}
where $\mathbf{\Theta}_{1:L}$ and $\mu$ are Lagrange multipliers.
The KKT conditions are
\begin{eqnarray}
 &  & \nabla_{\mathbf{\Sigma}_{l}}L\nonumber \\
 & = & w_{l}\mathbf{H}_{l,l}^{\dagger}\left(\mathbf{\Omega}_{l}+\mathbf{H}_{l,l}\mathbf{\Sigma}_{l}\mathbf{H}_{l,l}^{\dagger}\right)^{-1}\mathbf{H}_{l,l}+\mathbf{\Theta}_{l}-\mu\mathbf{I}\nonumber \\
 &  & -\sum_{k\ne l}w_{k}\mathbf{H}_{k,l}^{\dagger}\left(\mathbf{\Omega}_{k}^{-1}-\left(\mathbf{\Omega}_{k}+\mathbf{H}_{k,k}\mathbf{\Sigma}_{k}\mathbf{H}_{k,k}^{\dagger}\right)^{-1}\right)\mathbf{H}_{k,l}\nonumber \\
 & = & \mathbf{0},\label{eq:KKT_forward_1}
\end{eqnarray}
\[
\mu\left(P_{\text{T}}-\sum_{l=1}^{L}\text{Tr}\left(\mathbf{\Sigma}_{l}\right)\right)=0,
\]
\[
\text{tr}\left(\mathbf{\Sigma}_{l}\mathbf{\Theta}_{l}\right)=0,
\]
\[
\mathbf{\Sigma}_{l}\text{, }\mathbf{\Theta}_{l}\succcurlyeq0\text{, }\ensuremath{\mu\ge}0.
\]
 At a stationary point of problem (\ref{eq:WSRM_TP}), the transmit
signal covariance matrices $\mathbf{\Sigma}_{1:L}$ have the polite
water-filling structure \cite{Liu_IT10s_Duality_BMAC}. Recall that
in a single user MIMO channel, the optimal $\mathbf{\Sigma}$ is a
water-filling over channel $\mathbf{H}$, i.e., the eigenvectors of
$\mathbf{\Sigma}$ are the right singular vectors of $\mathbf{H}$
and the eigenvalues are calculated using water-filling of parallel
channels with singular values of $\mathbf{H}$ as channel gains. The
\emph{polite water-filling} structure is that the equivalent transmit
covariance matrix $\hat{\mathbf{\Omega}}_{l}^{\frac{1}{2}}\mathbf{\Sigma}_{l}\hat{\mathbf{\Omega}}_{l}^{\frac{1}{2}}$
is a water-filling over the equivalent post- and pre-whitened channel
$\bar{\mathbf{H}}_{l}=\mathbf{\Omega}_{l}^{-\frac{1}{2}}\mathbf{H}_{l,l}\hat{\mathbf{\Omega}}_{l}^{-\frac{1}{2}}$,
where the reverse link interference plus noise covariance $\hat{\mathbf{\Omega}}_{l}$
is calculated from $\hat{\mathbf{\Sigma}}_{1:L}$, and $\hat{\mathbf{\Sigma}}_{1:L}$
are calculated from $\mathbf{\Sigma}_{1:L}$ using the above mentioned
covariance transformation. The $\hat{\mathbf{\Sigma}}_{1:L}$ also
have the polite water-filling structure and are the stationary point
of the reverse link optimization problem (\ref{eq:WSRM_DTP}). In
the case of parallel channels, the polite water-filling will reduce
to the traditional water-filling. In MAC/BC, polite water-filling
structure is the optimal transmit signal structure for the capacity
region boundary points.

\subsubsection*{Polite Water-filling Algorithm}

The polite water-filling structure naturally suggests the iterative
\emph{polite water-filling algorithm}, Algorithm PP, in \cite{Liu_IT10s_Duality_BMAC}.
It works as follows. After initializing the reverse link interference
plus noise covariance matrices $\hat{\mathbf{\Omega}}_{1:L}$, we
perform a forward link polite water-filling to obtain $\mathbf{\Sigma}_{1:L}$.
The reverse link polite water-filling is performed to obtain $\hat{\mathbf{\Sigma}}_{1:L}$.
This finishes one iteration. The iterations stop when the change of
the objective function is less than a threshold or when a predetermined
number of iterations is reached. Because the algorithm enforces the
optimal signal structure at each iteration, it converges very fast
if it converges. In particular, for parallel channels, it gives the
optimal solution in half an iteration with initial values $\hat{\mathbf{\Omega}}_{l}=\mathbf{I},\ \forall l$.
Unfortunately, this algorithm is not guaranteed to converge, especially
in very strong interference cases.

\subsubsection*{Dual Transformation}

The following relations between $\mathbf{\Sigma}_{1:L}$ and $\hat{\mathbf{\Sigma}}_{1:L}$
at stationary points are proved using the polite water-filling structure
in \cite{Liu_IT10s_Duality_BMAC}. We name them \emph{dual transformation}
in this paper:
\begin{equation}
\hat{\mathbf{\Sigma}}_{l}=\frac{w_{l}}{\mu}\left(\mathbf{\Omega}_{l}^{-1}-\left(\mathbf{\Omega}_{l}+\mathbf{H}_{l,l}\mathbf{\Sigma}_{l}\mathbf{H}_{l,l}^{\dagger}\right)^{-1}\right)\text{, }l=1,\ldots,L;\label{eq:PWFS_forward}
\end{equation}
\begin{equation}
\mathbf{\Sigma}_{l}=\frac{w_{l}}{\hat{\mu}}\left(\hat{\mathbf{\Omega}}_{l}^{-1}-\left(\hat{\mathbf{\Omega}}_{l}+\mathbf{H}_{l,l}^{\dagger}\hat{\mathbf{\Sigma}}_{l}\mathbf{H}_{l,l}\right)^{-1}\right)\text{, }l=1,\ldots,L,\label{eq:PWFS_reverse}
\end{equation}
where the Lagrange multipliers $\mu$ and $\hat{\mu}$ are the Lagrange
multipliers of the forward and reverse links for the power constraints.
Equation (\ref{eq:PWFS_forward}) can be substituted into the KKT
condition (\ref{eq:KKT_forward_1}) to recover the \textit{\emph{polite
water-filling solution to the KKT conditions. In past works, the term
$\frac{w_{l}}{\mu}\left(\mathbf{\Omega}_{l}^{-1}-\left(\mathbf{\Omega}_{l}+\mathbf{H}_{l,l}\mathbf{\Sigma}_{l}\mathbf{H}_{l,l}^{\dagger}\right)^{-1}\right)$
in the KKT condition has always been the obstacle to an elegant solution.
Now we know it equals to $\hat{\mathbf{\Sigma}}_{l}$ at a stationary
point. The dual transformation is used in the next section to design
a new convergent algorithm. }}

\section{The Dual Link Algorithm\label{sec:DL_alg}}

\subsection{The Algorithm}

We propose a new algorithm, named Dual Link Algorithm, for the weighted
sum-rate maximization problem (\ref{eq:WSRM_TP}). It has fast and
monotonic convergence. The main idea is that, since we already know
the optimal input covariance matrices $\mathbf{\Sigma}_{1:L}$ and
$\hat{\mathbf{\Sigma}}_{1:L}$ must satisfy the dual transformation
(\ref{eq:PWFS_forward}) and (\ref{eq:PWFS_reverse}), we can directly
use these the dual transformation to update $\hat{\mathbf{\Sigma}}_{1:L}$
and $\mathbf{\Sigma}_{1:L}$, instead of solving the KKT conditions
and enforce the polite water-filling structure of $\hat{\mathbf{\Sigma}}_{1:L}$
and $\mathbf{\Sigma}_{1:L}$ as in the polite water-filling algorithms
\cite{Liu_IT10s_Duality_BMAC}. 

It is well known that equality $\sum_{l=1}^{L}\text{Tr}\left(\mathbf{\Sigma}_{l}\right)=P_{\text{T}}$
holds when $\mathbf{\Sigma}_{1:L}$ is a stationary point of problem
(\ref{eq:WSRM_TP}), e.g., \cite[Theorem 8 (item 3)]{Liu_10sTSP_Fairness_rate_polit_WF}.
This is because of the nonzero noise variance. It indicates that the
full power should always be used. Since the covariance transformation
\cite[Lemma 8]{Liu_IT10s_Duality_BMAC} preserves total power, we
also have $\sum_{l=1}^{L}\text{Tr}\left(\hat{\mathbf{\Sigma}}_{l}\right)=P_{\text{T}}$.
The Lagrange multipliers $\mu$ and $\hat{\mu}$should be chosen to
satisfy the power constraint $\sum_{l=1}^{L}\text{Tr}\left(\mathbf{\Sigma}_{l}\right)=P_{\text{T}}$
as
\begin{align}
\mu & =\frac{1}{P_{\text{T}}}\sum_{l=1}^{L}w_{l}\text{tr}\left(\mathbf{\Omega}_{l}^{-1}-\left(\mathbf{\Omega}_{l}+\mathbf{H}_{l,l}\mathbf{\Sigma}_{l}\mathbf{H}_{l,l}^{\dagger}\right)^{-1}\right)\label{eq:6}
\end{align}
\begin{align}
\hat{\mu} & =\frac{1}{P_{\text{T}}}\sum_{l=1}^{L}w_{l}\text{tr}\left(\hat{\mathbf{\Omega}}_{l}^{-1}-\left(\hat{\mathbf{\Omega}}_{l}+\mathbf{H}_{l,l}^{\dagger}\hat{\mathbf{\Sigma}}_{l}\mathbf{H}_{l,l}\right)^{-1}\right)\label{eq:hat_mu_value}
\end{align}

The above suggests the Dual Link Algorithm in Table Algorithm \ref{alg:1}
that takes advantage of the structure of the weighted sum-rate maximization
problem. A node who knows global channel state information runs the
algorithm. The algorithm starts by initializing $\mathbf{\Sigma}_{l}$'s
as random matrices or scaled identity matrices, which can be used
to calculate forward link interference plus noise covariance $\mathbf{\Omega}_{l}$'s.
Then, $\hat{\mathbf{\Sigma}}_{l}$'s of the virtual reverse links
can be calculated by the dual transformation (\ref{eq:PWFS_forward})
with $\mu$ given in (\ref{eq:6}). These $\hat{\mathbf{\Sigma}}_{l}$'s
are used to calculate virtual reverse link interference plus noise
covariance matrices $\hat{\mathbf{\Omega}}_{l}$'s. Then, $\mathbf{\Sigma}_{l}$'s
of the forward links can be calculated by the dual transformation
(\ref{eq:PWFS_reverse}) with $\hat{\mu}$ given in (\ref{eq:hat_mu_value}).
The above is repeated until the weighted sum rate converges or a fixed
number of iterations are reached.

The most important properties of the dual link algorithm is that,
unlike other algorithms for this problem, it is ideally suited for
\emph{distributed} implementation and is \emph{scalable} to network
size. This will be discussed briefly in Section \ref{sec:Distributed-Implementation}. 

As confirmed by the proof and numerical experiments, Dual Link Algorithm
has monotonic convergence and is almost as fast as the polite water-filling
(PWF) algorithm. It converges to a stationary point of both problem
(\ref{eq:WSRM_TP}) and its dual (\ref{eq:WSRM_DTP}) simultaneously,
and both (\ref{eq:PWFS_forward}) and (\ref{eq:PWFS_reverse}) achieve
the same sum-rate at the stationary point. 

\begin{algorithm}
\begin{raggedright}
1. Initialize $\mathbf{\Sigma}_{l}$'s, s.t. $\sum_{l=1}^{L}\text{Tr}\left(\mathbf{\Sigma}_{l}\right)=P_{\text{T}}$
\par\end{raggedright}

\begin{raggedright}
2. $R\Leftarrow\sum_{l=1}^{L}w_{l}\mathcal{I}_{l}\left(\mathbf{\Sigma}_{1:L}\right)$
\par\end{raggedright}

\begin{raggedright}
3. Repeat
\par\end{raggedright}

\begin{raggedright}
4. $R^{'}\Leftarrow R$
\par\end{raggedright}

\begin{raggedright}
5. $\mathbf{\Omega}_{l}\Leftarrow\mathbf{I}+\sum_{k\ne l}\mathbf{H}_{l,k}\mathbf{\Sigma}_{k}\mathbf{H}_{l,k}^{\dagger}$
\par\end{raggedright}

\begin{raggedright}
6. $\hat{\mathbf{\Sigma}}_{l}\Leftarrow\frac{P_{\text{T}}w_{l}\left(\mathbf{\Omega}_{l}^{-1}-\left(\mathbf{\Omega}_{l}+\mathbf{H}_{l,l}\mathbf{\Sigma}_{l}\mathbf{H}_{l,l}^{\dagger}\right)^{-1}\right)}{\sum_{l=1}^{L}w_{l}\text{tr}\left(\mathbf{\Omega}_{l}^{-1}-\left(\mathbf{\Omega}_{l}+\mathbf{H}_{l,l}\mathbf{\Sigma}_{l}\mathbf{H}_{l,l}^{\dagger}\right)^{-1}\right)}$
\par\end{raggedright}

\begin{raggedright}
7. $\hat{\mathbf{\Omega}}_{l}\Leftarrow\mathbf{I}+\sum_{k\ne l}\mathbf{H}_{k,l}^{\dagger}\hat{\mathbf{\Sigma}}_{k}\mathbf{H}_{k,l}$
\par\end{raggedright}

\begin{raggedright}
8. $\mathbf{\Sigma}_{l}=\frac{P_{\text{T}}w_{l}\left(\hat{\mathbf{\Omega}}_{l}^{-1}-\left(\hat{\mathbf{\Omega}}_{l}+\mathbf{H}_{l,l}^{\dagger}\hat{\mathbf{\Sigma}}_{l}\mathbf{H}_{l,l}\right)^{-1}\right)}{\sum_{l=1}^{L}w_{l}\text{tr}\left(\hat{\mathbf{\Omega}}_{l}^{-1}-\left(\hat{\mathbf{\Omega}}_{l}+\mathbf{H}_{l,l}^{\dagger}\hat{\mathbf{\Sigma}}_{l}\mathbf{H}_{l,l}\right)^{-1}\right)}$
\par\end{raggedright}

\begin{raggedright}
9. $R\Leftarrow\sum_{l=1}^{L}w_{l}\mathcal{I}_{l}\left(\mathbf{\Sigma}_{1:L}\right)$
\par\end{raggedright}

\begin{raggedright}
10. until $\left|R-R^{'}\right|\le\epsilon$ or a fixed number of
iterations are reached. 
\par\end{raggedright}

\caption{\label{alg:1}Dual Link Algorithm}
\end{algorithm}

\subsection{Preliminaries of the Convergence Proof}

In the following sections, we prove the monotonic convergence of Algorithm
\ref{alg:1}. As will be seen later, the proof uses only very general
convex analysis, as well as a particular scaling invariance property
that we identify for the weighted sum-rate maximization problem. We
expect that the scaling invariance holds for and our proof technique
applies to many non-convex problems in communication networks that
involve the rate or throughput maximization.

\subsubsection{Equivalent Problem and the Lagrange Function}

The weighted sum-rate maximization problem (\ref{eq:WSRM_TP}) is
equivalent to the following problem by considering the interference
plus noise covariance matrices as additional variables with additional
equality constraints:
\begin{eqnarray}
\max_{\Sigma_{1:L},\Omega_{1:L}} &  & \sum_{l=1}^{L}w_{l}\left(\log\left|\mathbf{\Omega}_{l}+\mathbf{H}_{l,l}\mathbf{\Sigma}_{l}\mathbf{H}_{l,l}^{\dagger}\right|-\log\left|\mathbf{\Omega}_{l}\right|\right)\nonumber \\
s.t. &  & \mathbf{\Sigma}_{l}\succeq0\text{, }\forall l,\nonumber \\
 &  & \sum_{l=1}^{L}\text{Tr}\left(\mathbf{\Sigma}_{l}\right)\le P_{\text{T}},\nonumber \\
 &  & \mathbf{\Omega}_{l}=\mathbf{I}+\sum_{k\neq l}\mathbf{H}_{l,k}\mathbf{\Sigma}_{k}\mathbf{H}_{l,k}^{\dagger},\forall l,\label{eq:WSRM_TPeq}
\end{eqnarray}
which is still non-convex. Consider the Lagrangian of the above problem
\begin{eqnarray*}
 &  & F(\mathbf{\Sigma},\mathbf{\Omega},\mathbf{\Lambda},\mu)\\
 & = & \sum_{l=1}^{L}w_{l}\left(\log\left|\mathbf{\Omega}_{l}+\mathbf{H}_{l,l}\mathbf{\Sigma}_{l}\mathbf{H}_{l,l}^{\dagger}\right|-\log\left|\mathbf{\Omega}_{l}\right|\right)\\
 &  & +\mu\{P_{\text{T}}-\sum_{l=1}^{L}\text{Tr}(\mathbf{\Sigma}_{l})\}\\
 &  & +\sum_{l=1}^{L}\text{Tr}\left(\mathbf{\Lambda}_{l}\left(\mathbf{\Omega}_{l}-\mathbf{I}-\sum_{k\neq l}\mathbf{H}_{l,k}\mathbf{\Sigma}_{k}\mathbf{H}_{l,k}^{\dagger}\right)\right),
\end{eqnarray*}
where $\mathbf{\Sigma}$ represents $\mathbf{\Sigma}_{1:L}$; $\mathbf{\Omega}$
represents $\mathbf{\Omega}_{1:L}$; $\mathbf{\Lambda}$ represents
$\mathbf{\Lambda}_{1:L}$; the domain of $F$ is $\{\mathbf{\Sigma},\mathbf{\Omega},\mathbf{\Lambda},\mu|\mathbf{\Sigma}_{l}\in\mathbb{H}_{+}^{L_{T_{l}}\times L_{T_{l}}},\mathbf{\Omega}_{l}\in\mathbb{H}_{++}^{L_{R_{l}}\times L_{R_{l}}},\mathbf{\Lambda}_{l}\in\mathbb{H}^{L_{R_{l}}\times L_{R_{l}}},\mu\in\mathbb{R}^{+},\forall l\}$.
Here $\mathbb{H}^{n\times n}$, $\mathbb{H}_{+}^{n\times n}$, and
$\mathbb{H}_{++}^{n\times n}$ are the sets of $n\times n$ Hermitian
matrices, positive semidefinite matrices, and positive definite matrices
respectively. 

One can easily verify that the function $F$ is concave in $\mathbf{\Sigma}$
and convex in $\mathbf{\Omega}$. Furthermore, the gradients are given
by
\begin{eqnarray*}
\nabla_{\mathbf{\Sigma}_{l}}F & = & w_{l}\mathbf{H}_{l,l}^{\dagger}\left(\mathbf{\Omega}_{l}+\mathbf{H}_{l,l}\mathbf{\Sigma}_{l}\mathbf{H}_{l,l}^{\dagger}\right)^{-1}\mathbf{H}_{l,l}\\
 &  & -\mu\mathbf{I}-\sum_{k\neq l}\mathbf{H}_{k,l}^{\dagger}\mathbf{\Lambda}_{l}\mathbf{H}_{k,l},\\
\nabla_{\mathbf{\Omega}_{l}}F & = & w_{l}\left(\left(\mathbf{\Omega}_{l}+\mathbf{H}_{l,l}\mathbf{\Sigma}_{l}\mathbf{H}_{l,l}^{\dagger}\right)^{-1}-\mathbf{\Omega}_{l}^{-1}\right)+\mathbf{\Lambda}_{l}.
\end{eqnarray*}

Now suppose that we have the pair $(\mathbf{\Sigma},\mathbf{\Omega})$
such that
\begin{eqnarray*}
\sum_{l=1}^{L}\text{Tr}(\mathbf{\Sigma}_{l}) & = & P_{\text{T}},\\
\mathbf{\Omega}_{l} & = & \mathbf{I}+\sum_{k\neq l}\mathbf{H}_{l,k}\mathbf{\Sigma}_{k}\mathbf{H}_{l,k}^{\dagger},
\end{eqnarray*}
then,
\begin{eqnarray*}
 &  & F(\mathbf{\Sigma}_{1:L},\mathbf{\Omega}_{1:L},\mathbf{\Lambda}_{1:L},\mu)\\
 & = & \sum_{l=1}^{L}w_{l}\left(\log\left|\Omega_{l}+\mathbf{H}_{l,l}\mathbf{\Sigma}_{l}\mathbf{H}_{l,l}^{\dagger}\right|-\log\left|\mathbf{\Omega}_{l}\right|\right),
\end{eqnarray*}
which is the original weighted sum-rate function. For notational simplicity,
denote the weighted sum-rate function by $V(\Sigma)$, i.e.,
\begin{eqnarray*}
V(\mathbf{\Sigma}) & = & \sum_{l=1}^{L}w_{l}\left(\log\left|\mathbf{I}+\sum_{k\neq l}\mathbf{H}_{l,k}\mathbf{\Sigma}_{k}\mathbf{H}_{l,k}^{\dagger}+\mathbf{H}_{l,l}\mathbf{\Sigma}_{l}\mathbf{H}_{l,l}^{\dagger}\right|\right.\\
 &  & -\left.\log\left|\mathbf{I}+\sum_{k\neq l}\mathbf{H}_{l,k}\mathbf{\Sigma}_{k}\mathbf{H}_{l,k}^{\dagger}\right|\right).
\end{eqnarray*}

\subsubsection{Solution of the first-order condition}

Suppose that we want to solve the following system of equations in
terms of $(\mathbf{\Sigma},\mathbf{\Omega})$ for given $(\mathbf{\Lambda},\mu)$:
\begin{eqnarray*}
\nabla_{\mathbf{\Sigma}_{l}}F & = & 0,
\end{eqnarray*}
\begin{eqnarray*}
\nabla_{\mathbf{\Omega}_{l}}F & = & 0.
\end{eqnarray*}
Define
\begin{eqnarray*}
\hat{\mathbf{\Sigma}}_{l} & = & \frac{1}{\mu}\mathbf{\Lambda}_{l},\\
\hat{\mathbf{\Omega}}_{l} & = & \mathbf{I}+\sum_{k\neq l}\mathbf{H}_{k,l}^{\dagger}\hat{\mathbf{\Sigma}}_{l}\mathbf{H}_{k,l},
\end{eqnarray*}
the above system of equations becomes
\begin{eqnarray}
\hat{\mathbf{\Sigma}}_{l} & = & \frac{w_{l}}{\mu}\left(\mathbf{\Omega}_{l}^{-1}-\left(\mathbf{\Omega}_{l}+\mathbf{H}_{l,l}\mathbf{\Sigma}_{l}\mathbf{H}_{l,l}^{\dagger}\right)^{-1}\right),\label{eq:1stOrdCon1}\\
\hat{\mathbf{\Omega}}_{l} & = & \frac{w_{l}}{\mu}\mathbf{H}_{l,l}^{\dagger}\left(\mathbf{\Omega}_{l}+\mathbf{H}_{l,l}\mathbf{\Sigma}_{l}\mathbf{H}_{l,l}^{\dagger}\right)^{-1}\mathbf{H}_{l,l}.\label{eq:1stOrdCon2}
\end{eqnarray}

An explicit solution to this system of equations is given by
\begin{eqnarray}
\mathbf{\Sigma}_{l} & = & \frac{w_{l}}{\mu}\left(\hat{\mathbf{\Omega}}_{l}^{-1}-\left(\hat{\mathbf{\Omega}}_{l}+\mathbf{H}_{l,l}^{\dagger}\hat{\mathbf{\Sigma}}_{l}\mathbf{H}_{l,l}\right)^{-1}\right)\label{eq:kktsol}\\
\mathbf{\Omega}_{l} & = & \frac{w_{l}}{\mu}\mathbf{H}_{l,l}\left(\mathbf{H}_{l,l}^{\dagger}\hat{\mathbf{\Sigma}}_{l}\mathbf{H}_{l,l}+\hat{\mathbf{\Omega}}_{l}\right)^{-1}\mathbf{H}_{l,l}^{\dagger}.\label{eq:kktsol2}
\end{eqnarray}
The detailed proof of this solution can be found in \cite{Yu_IT06_Minimax_duality,ChenL_CoRR2013_WSR_MIMO_minimax}.
\begin{rem}
(\ref{eq:kktsol}) and (\ref{eq:kktsol2}) are actually the first-order
optimality conditions of (\ref{eq:WSRM_TPeq})'s dual problem which
is equivalent to (\ref{eq:WSRM_DTP}). Algorithm \ref{alg:1} uses
(\ref{eq:1stOrdCon1}) and (\ref{eq:kktsol}) to update $\hat{\mathbf{\Sigma}}_{1:L}$
and $\mathbf{\Sigma}_{1:L}$. When it converges, equations (\ref{eq:1stOrdCon1})-(\ref{eq:kktsol2})
will all hold, and the KKT conditions of problem (\ref{eq:WSRM_TPeq})
and its dual will all be satisfied.
\end{rem}

\subsection{Convergence Results}

We are ready to present the following two main convergence results
regarding Algorithm \ref{alg:1}. Denote by $\mathbf{{\Sigma}}^{(n)}$
the $\mathbf{\Sigma}$ value at the $n$-th iteration of Algorithm
\ref{alg:1}. 
\begin{thm}
\label{thm:monotone}The objective value, i.e., the weighted sum-rate,
is monotonically increasing in Algorithm \ref{alg:1}, ${\it {i.e.}}$,
\begin{eqnarray*}
V(\mathbf{{\Sigma}}^{(n)}) & \leq & V(\mathbf{{\Sigma}}^{(n+1)}).
\end{eqnarray*}

\end{thm}
From the above theorem, the following conclusion is immediate.
\begin{cor}
\label{cor:limitpoint}The sequence $V_{n}=V(\mathbf{{\Sigma}}^{(n)})$\textup{
converges to some limit point $V_{\infty}$.}\end{cor}
\begin{IEEEproof}
Since $V(\mathbf{{\Sigma}})$ is a continuous function and its domain
$\{\mathbf{{\Sigma}}|\mathbf{{\Sigma}}_{l}\succeq\mathbf{0},\text{{Tr}}(\mathbf{{\Sigma}})\leqq P_{\text{T}},\forall l\}$
is a compact set, $V_{n}$ is bounded above. From Theorem \ref{thm:monotone},
$\{V_{n}\}$ is a monotone increasing sequence, therefore there exists
a limit point $V_{\infty}$ such that $\lim_{n\rightarrow\infty}V_{n}=V_{\infty}$.
\end{IEEEproof}
If we define a stationary point ($\mathbf{{\Sigma}}_{L}^{\star})$
of Algorithm \ref{alg:1}, $\mathbf{{\Sigma}}^{(n)}=\mathbf{{\Sigma}}^{\star}$
implies $\mathbf{{\Sigma}}^{(n+k)}=\mathbf{{\Sigma}}^{\star}$ for
all $k=0,1,\cdots$, then we have the following result.
\begin{thm}
\label{thm:convergence} Algorithm \ref{alg:1} converges to a stationary
point $\mathbf{{\Sigma}}_{1:L}^{\star}$\textup{.}
\end{thm}
The above implies that both the weighted sum rate and the transmit
signal covariance matrices converge. The proof of Theorems \ref{thm:monotone}
and \ref{thm:convergence} will be presented later in this section.
Before that, we first establish a few inequalities and identify a
particular scaling property of the Lagrangian $F$.

\subsubsection{The first inequality}

Suppose that we have a feasible point $\mathbf{\Sigma}^{(n)}\succeq0$,
and
\begin{eqnarray}
\sum_{l=1}^{L}\text{Tr}\left(\mathbf{\Sigma}_{l}^{(n)}\right) & = & P_{\text{T}}.\label{eq:1edueto1}
\end{eqnarray}
 In Algorithm \ref{alg:1}, we generate $\mathbf{\Omega}_{l}^{(n)}$
such that
\begin{eqnarray}
\mathbf{\Omega}_{l}^{(n)} & = & \mathbf{I}+\sum_{k\neq l}\mathbf{H}_{l,k}\mathbf{\Sigma}_{k}^{(n)}\mathbf{H}_{l,k}^{\dagger}.\label{eq:1edueto2}
\end{eqnarray}
Now we have a pair $(\mathbf{\Sigma}^{(n)},\mathbf{\Omega}^{(n)})$.
Using this pair, we can compute $(\mathbf{\Lambda}_{1:L}^{(n)},\mu^{(n)})$
as
\begin{eqnarray*}
\mathbf{\Lambda}_{l}^{(n)} & = & w_{l}\left({\mathbf{\Omega}_{l}^{(n)}}^{-1}-\left(\mathbf{\Omega}_{l}^{(n)}+\mathbf{H}_{l,l}\mathbf{\Sigma}_{l}^{(n)}\mathbf{H}_{l,l}^{\dagger}\right)^{-1}\right),\\
\mu^{(n)} & = & \frac{1}{P_{\text{T}}}\sum_{l=1}^{L}\text{Tr}\left(\mathbf{\Lambda}_{l}^{(n)}\right).
\end{eqnarray*}
Note that $\hat{\Sigma}_{l}^{(n)}$ in Algorithm \ref{alg:1} is equal
to
\begin{eqnarray*}
\hat{\Sigma}_{l}^{(n)} & = & \frac{{\mathbf{\Lambda}}_{l}^{(n)}}{\mu^{(n)}}.
\end{eqnarray*}
From this and (\ref{eq:1stOrdCon1}), the gradient of $F$ with respect
to $\mathbf{\Omega}$ at the point $(\mathbf{\Sigma}^{(n)},\mathbf{\Omega}^{(n)})$
vanishes, i.e.,
\begin{eqnarray*}
\nabla_{\mathbf{\Omega}}F(\mathbf{\Sigma}^{(n)},\mathbf{\Omega},\mathbf{\Lambda}^{(n)},\mu^{(n)})|_{\mathbf{\Omega}^{(n)}} & = & 0.
\end{eqnarray*}
Since $F$ is convex in $\mathbf{\Omega}$, if we fix $\mathbf{\Sigma}=\mathbf{\Sigma}^{(n)}$,
then $\mathbf{\Omega}^{(n)}$ is a global minimizer of $F$. In other
words,
\begin{equation}
F(\mathbf{\Sigma}^{(n)},\mathbf{\Omega}^{(n)},\mathbf{\Lambda}^{(n)},\mu^{(n)})\leq F(\mathbf{\Sigma}^{(n)},\mathbf{\Omega},\mathbf{\Lambda}^{(n)},\mu^{(n)})\label{eq:firstineq}
\end{equation}
for all $\mathbf{\Omega}\succ0$. %Scailing property

\subsubsection{Scaling invariance of $F$}

We will identify a remarkable scaling invariance property of $F$,
which plays a key role in the convergence proof of Algorithm \ref{alg:1}.
For given $(\mathbf{\Sigma}^{(n)},\mathbf{\Omega}^{(n)},\mathbf{\Lambda}^{(n)},\mu^{(n)})$,
we have
\begin{eqnarray}
 &  & F(\frac{1}{\alpha}\mathbf{\Sigma}^{(n)},\frac{1}{\alpha}\mathbf{\Omega}^{(n)},\alpha\mathbf{\Lambda}^{(n)},\alpha\mu^{(n)})\nonumber \\
 & = & F(\mathbf{\Sigma}^{(n)},\mathbf{\Omega}^{(n)},\mathbf{\Lambda}^{(n)},\mu^{(n)})\label{eq:scaling}
\end{eqnarray}
 for all $\alpha>$0. To show this scaling invariance property, note
that
\begin{eqnarray*}
\mathbf{\Omega}_{l}^{(n)}-\sum_{k\neq l}\mathbf{H}_{l,k}\mathbf{\Sigma}_{k}^{(n)}\mathbf{H}_{l,k}^{\dagger} & = & \mathbf{I},\\
\sum_{l=1}^{L}\text{Tr}(\mathbf{\Sigma}_{l}^{(n)}) & = & P_{\text{T}},\\
P_{\text{T}}\mu^{(n)} & = & \sum_{l=1}^{L}\text{Tr}(\mathbf{\Lambda}_{l}^{(n)}).
\end{eqnarray*}
Applying the above equalities and some mathematical manipulations,
we have
\begin{eqnarray*}
 &  & F(\frac{1}{\alpha}\mathbf{\Sigma}^{(n)},\frac{1}{\alpha}\mathbf{\Omega}^{(n)},\alpha\mathbf{\Lambda}^{(n)},\alpha\mu^{(n)})\\
 & = & \sum_{l=1}^{L}w_{l}\left(\log\left|\mathbf{\Omega}_{l}^{(n)}+\mathbf{H}_{l,l}\mathbf{\Sigma}_{l}^{(n)}\mathbf{H}_{l,l}^{\dagger}\right|-\log\left|\mathbf{\Omega}_{l}^{(n)}\right|\right)\\
 &  & +\alpha\mu^{(n)}\{P_{\text{T}}-\frac{1}{\alpha}P_{\text{T}}\}+\sum_{l=1}^{L}\text{Tr}\left(\alpha\mathbf{\Lambda}_{l}^{(n)}\left(\frac{1}{\alpha}\mathbf{I}-\mathbf{I}\right)\right)\\
 & = & \sum_{l=1}^{L}w_{l}\left(\log\left|\mathbf{\Omega}_{l}^{(n)}+\mathbf{H}_{l,l}\mathbf{\Sigma}_{l}^{(n)}\mathbf{H}_{l,l}^{\dagger}\right|-\log\left|\mathbf{\Omega}_{l}^{(n)}\right|\right)\\
 &  & +(\alpha-1)\mu^{(n)}P_{\text{T}}+(1-\alpha)\sum_{l=1}^{L}\text{Tr}(\mathbf{\Lambda}_{l}^{(n)})\\
 & = & \sum_{l=1}^{L}w_{l}\left(\log\left|\mathbf{\Omega}_{l}^{(n)}+\mathbf{H}_{l,l}\mathbf{\Sigma}_{l}^{(n)}\mathbf{H}_{l,l}^{\dagger}\right|-\log\left|\mathbf{\Omega}_{l}^{(n)}\right|\right)\\
 & = & F(\mathbf{\Sigma}^{(n)},\mathbf{\Omega}^{(n)},\mathbf{\Lambda}^{(n)},\mu^{(n)}),
\end{eqnarray*}
where the first equality uses the fact that
\begin{eqnarray*}
 &  & \log\left|\frac{1}{\alpha}\left(\mathbf{\Omega}_{l}^{(n)}+\mathbf{H}_{l,l}\mathbf{\Sigma}_{l}^{(n)}\mathbf{H}_{l,l}^{\dagger}\right)\right|-\log\left|\frac{1}{\alpha}\mathbf{\Omega}_{l}^{(n)}\right|\\
 & = & \log\left|\mathbf{\Omega}_{l}^{(n)}+\mathbf{H}_{l,l}\mathbf{\Sigma}_{l}^{(n)}\mathbf{H}_{l,l}^{\dagger}\right|-\log\left|\mathbf{\Omega}_{l}^{(n)}\right|.
\end{eqnarray*}

Furthermore,
\begin{eqnarray}
 &  & \nabla_{\mathbf{\Omega}_{l}}F(\frac{1}{\alpha}\mathbf{\Sigma}^{(n)},\mathbf{\Omega},\alpha\mathbf{\Lambda}^{(n)},\alpha\mu^{(n)})|_{\frac{1}{\alpha}\mathbf{\Omega}^{(n)}}\nonumber \\
 & = & w_{l}\left(\left(\frac{1}{\alpha}\mathbf{\Omega}_{l}^{(n)}+\mathbf{H}_{l,l}\frac{1}{\alpha}\mathbf{\Sigma}_{l}^{(n)}\mathbf{H}_{l,l}^{\dagger}\right)^{-1}-\left(\frac{1}{\alpha}\mathbf{\Omega}_{l}^{(n)}\right)^{-1}\right)\nonumber \\
 &  & +\alpha\mathbf{\Lambda}_{l}^{(n)}\nonumber \\
 & = & \alpha\nabla_{\mathbf{\Omega}_{l}}F(\mathbf{\Sigma}^{(n)},\mathbf{\Omega},\mathbf{\Lambda}^{(n)},\mu^{(n)})|_{\mathbf{\Omega}^{(n)}}\nonumber \\
 & = & 0,\ \forall l.\label{eq:scaling-gradient}
\end{eqnarray}
Therefore, $\frac{1}{\alpha}\mathbf{\Omega}^{(n)}$ is a global minimizer
of $F(\frac{1}{\alpha}\mathbf{\Sigma}^{(n)},\mathbf{\Omega},\alpha\mathbf{\Lambda}^{(n)},\alpha\mu^{(n)})$,
as $F$ is convex in $\Omega$.%Primal point 

\subsubsection{The second and third inequalities}

Given $(\alpha\mathbf{\Lambda}^{(n)},\alpha\mu^{(n)})$, we generate
$\tilde{{\mathbf{{\Sigma}}}},\tilde{{\mathbf{{\Omega}}}}$ using equation
(\ref{eq:kktsol}) and (\ref{eq:kktsol2}). If we choose $\alpha$
so that
\begin{eqnarray}
\sum_{l=1}^{L}\text{Tr(}\tilde{{\mathbf{{\Sigma}}}_{l}}) & = & P_{\text{T}},\label{eq:3edueto1}
\end{eqnarray}
 then $\tilde{{\mathbf{{\Sigma}}}}=\mathbf{\Sigma}^{(n+1)}$ in Algorithm
\ref{alg:1}. Since $(\mathbf{\Sigma}^{(n+1)},\tilde{\mathbf{{\Omega}}})$
is chosen to make the gradients zero: 
\begin{eqnarray*}
\nabla_{\mathbf{\Sigma}}F(\mathbf{\Sigma},\mathbf{\tilde{{\Omega}}},\alpha\mathbf{\Lambda}^{(n)},\alpha\mu^{(n)})|_{\mathbf{\Sigma}^{(n+1)}} & = & 0,\\
\nabla_{\mathbf{\Omega}}F(\mathbf{\Sigma}^{(n+1)},\mathbf{\Omega},\alpha\mathbf{\Lambda}^{(n)},\alpha\mu^{(n)})|_{\tilde{{\mathbf{\Omega}}}} & = & 0,
\end{eqnarray*}
we conclude that $\mathbf{\Sigma}^{(n+1)}$ is a global maximizer,
i.e.,
\begin{equation}
F(\mathbf{\Sigma},\tilde{{\mathbf{\Omega}}},\alpha\mathbf{\Lambda}^{(n+1)},\alpha\mu^{(n+1)})\leq F(\mathbf{\Sigma}^{(n+1)},\tilde{{\mathbf{\Omega}}},\alpha\mathbf{\Lambda}^{(n)},\alpha\mu^{(n)})\label{eq:secondineq}
\end{equation}
for all $\mathbf{\Sigma}\succeq0$; and $\tilde{{\mathbf{\Omega}}}$
is a global minimizer, i.e.,
\begin{equation}
F(\mathbf{\Sigma}^{(n+1)},\tilde{{\mathbf{\Omega}}},\alpha\mathbf{\Lambda}^{(n)},\alpha\mu^{(n)})\leq F(\mathbf{\Sigma}^{(n+1)},\mathbf{\Omega},\alpha\mathbf{\Lambda}^{(n)},\alpha\mu^{(n)})\label{eq:thridineq}
\end{equation}
for all $\mathbf{\Omega}\succ0$. %Chain of inequalities

\subsubsection{Proof of Theorem \ref{thm:monotone}}

With the three inequalities (\ref{eq:firstineq}, \ref{eq:secondineq},
\ref{eq:thridineq}) obtained above, we are ready to prove Theorem
\ref{thm:monotone}. As in Algorithm \ref{alg:1}
\begin{eqnarray}
\mathbf{{\Omega}}_{l}^{(n+1)} & = & \mathbf{I}+\sum_{k\neq l}\mathbf{H}_{l,k}\mathbf{\Sigma}_{k}^{(n+1)}\mathbf{H}_{l,k}^{\dagger},\label{eq:3edueto2}
\end{eqnarray}
 we have
\begin{eqnarray}
 &  & V(\mathbf{\Sigma}^{(n)})\nonumber \\
 & = & F(\mathbf{\Sigma}^{(n)},\mathbf{\Omega}^{(n)},\mathbf{\Lambda}^{(n)},\mu^{(n)})\label{eq:1e}\\
 & = & F(\frac{1}{\alpha}\mathbf{\Sigma}^{(n)},\frac{1}{\alpha}\mathbf{\Omega}^{(n)},\alpha\mathbf{\Lambda}^{(n)},\alpha\mu^{(n)})\label{eq:2e}\\
 & \le & F(\frac{1}{\alpha}\mathbf{\Sigma}^{(n)},\tilde{{\mathbf{{\Omega}}}},\alpha\mathbf{\Lambda}^{(n)},\alpha\mu^{(n)})\label{eq:1ine}\\
 & \le & F(\mathbf{\Sigma}^{(n+1)},\tilde{{\mathbf{{\Omega}}}},\alpha\mathbf{\Lambda}^{(n)},\alpha\mu^{(n)})\label{eq:2ine}\\
 & \le & F(\mathbf{\Sigma}^{(n+1)},\mathbf{{\Omega}}^{(n+1)},\alpha\mathbf{\Lambda}^{(n)},\alpha\mu^{(n)})\label{eq:3ine}\\
 & = & V(\mathbf{\Sigma}^{(n+1)}),\label{eq:3e}
\end{eqnarray}
where (\ref{eq:1e}) follows from the satisfied constraints (\ref{eq:1edueto1},
\ref{eq:1edueto2}); (\ref{eq:2e}) follows from the scaling invariance
(\ref{eq:scaling}); (\ref{eq:1ine}) follows from convexity and scaling
invariance (\ref{eq:firstineq}, \ref{eq:scaling-gradient}); (\ref{eq:2ine})
follows from the second inequality (\ref{eq:secondineq}); (\ref{eq:3ine})
follows from the third inequality (\ref{eq:thridineq}); (\ref{eq:3e})
follows from the satisfied constraints (\ref{eq:3edueto1}, \ref{eq:3edueto2}).

\subsubsection{Proof of Theorem \ref{thm:convergence}}

We have shown in Corollary \ref{cor:limitpoint} that $V_{n}$ converges
to a limit point under Algorithm \ref{alg:1}. To show the convergence
of the algorithm, it is enough to show that if $V(\mathbf{\Sigma}^{(n)})=V(\mathbf{\Sigma}^{(n+1)})$,
then $\mathbf{\Sigma}^{(n+1)}=\mathbf{\Sigma}^{(n+k)}$ for all $k=1,2,\cdots$.
Suppose $V(\mathbf{\Sigma}^{(n)})=V(\mathbf{\Sigma}^{(n+1)})$, then
from the proof in the above, we have
\begin{eqnarray*}
 &  & F(\mathbf{\Sigma}^{(n+1)},\mathbf{{\Omega}}^{(n+1)},\alpha\mathbf{\Lambda}^{(n)},\alpha\mu^{(n)})\\
 & = & F(\mathbf{\Sigma}^{(n+1)},\tilde{{\mathbf{{\Omega}}}},\alpha\mathbf{\Lambda}^{(n)},\alpha\mu^{(n)}).
\end{eqnarray*}
Since $\tilde{{\mathbf{{\Omega}}}}$ is a global minimizer, the above
equality implies $\mathbf{{\Omega}}^{(n+1)}$ is a global minimizer
too. From the first order condition for optimality, we have
\begin{eqnarray*}
 &  & \nabla_{\mathbf{\Omega}_{l}}F(\mathbf{\Sigma}^{(n+1)},\mathbf{\Omega},\alpha\mathbf{\Lambda}^{(n)},\alpha\mu^{(n+1)})|_{\mathbf{{\Omega}}^{(n+1)}}\\
 & = & w_{l}\left(\left(\mathbf{{\Omega}}_{l}^{(n+1)}+\mathbf{H}_{l,l}\mathbf{\Sigma}_{l}^{(n+1)}\mathbf{H}_{l,l}^{\dagger}\right)^{-1}-\{\mathbf{{\Omega}}_{l}^{(n+1)}\}^{-1}\right)\\
 &  & +\alpha\mathbf{\Lambda}_{l}^{(n)}\\
 & = & 0.
\end{eqnarray*}

On the other hand, we generate $\mathbf{\Lambda}^{(n+1)}$ such that
\begin{eqnarray*}
 &  & \mathbf{\Lambda}_{l}^{(n+1)}\\
 & = & w_{l}\left({\mathbf{\Omega}_{l}^{(n+1)}}^{-1}-\left(\mathbf{\Omega}_{l}^{(n+1)}+\mathbf{H}_{l,l}\mathbf{\Sigma}_{l}^{(n+1)}\mathbf{H}_{l,l}^{\dagger}\right)^{-1}\right)\\
 & = & \alpha\mathbf{\Lambda}_{l}^{(n)}.
\end{eqnarray*}
This shows $\hat{\mathbf{\Sigma}}^{(n+1)}\propto\hat{\mathbf{\Sigma}}^{(n)}$.
However, since the trace of each matrix is same, we conclude that
\begin{eqnarray*}
\hat{\mathbf{\Sigma}}^{(n+1)} & = & \hat{\mathbf{\Sigma}}^{(n)}.
\end{eqnarray*}
 From this it is obvious that $\hat{\mathbf{\Sigma}}^{(n)}=\hat{\mathbf{\Sigma}}^{(n+1)}=\cdots$
and $\mathbf{\Sigma}^{(n+1)}=\mathbf{\Sigma}^{(n+2)}=\cdots$.
\begin{rem}
When the algorithm converges, the pair $(\mathbf{\Sigma},\mathbf{\Omega})$
satisfies the first order optimality condition for $F$. Moreover,
since $\sum_{l=1}^{L}\text{Tr}\left(\mathbf{\Sigma}_{l}\right)=P_{\text{T}}$,
and $\mathbf{\Omega}_{l}=\mathbf{I}+\sum_{k\neq l}\mathbf{H}_{l,k}\mathbf{\Sigma}_{k}\mathbf{H}_{l,k}^{\dagger},\forall l$,
$(\alpha\mathbf{\Lambda},\alpha\mu)$ also satisfies the first order
optimality condition for $F$. This implies that the pair $(\mathbf{\Sigma},\mathbf{\Omega},\alpha\mathbf{\Lambda},\alpha\mu)$
is a saddle point of $F$, which means that they are indeed a primal-dual
point that solves the KKT system of the weighted sum-rate maximization.
\end{rem}

\section{Extensions}

In this section, we extend the dual link algorithm for the cases of
colored noise and weighted sum power constraint. We also discuss the
dual link algorithm's advantages in distributed implementation.

\subsection{Colored Noise and Weighted Power Constraint\label{sec:Colored-Noise}}

In practice, the noise, including thermal noise and interference from
other non-cooperating networks, may not be white as assumed in (\ref{eq:Omega}).
Due to nonuniform devices, we may use weighted sum power constraint.
The dual link algorithm can be easily adjusted to solve the weighted
sum-rate maximization problem with colored noise and a weighted power
constraint. The key idea is to use proper pre- and post-whitening
method to convert the problem back to the white noise and sum power
constraint form, while keeping the reciprocity property of the forward
and reverse link channels so that the original dual link algorithm
can be readily applied.

We use the following equivalence to find the proper whitening algorithm
\cite{Liu_IT10s_Duality_BMAC}. Assume the noise covariance matrix
of forward link $l$ is a positive definite matrix $\mathbf{W}_{l}\in\mathbb{C}^{L_{R_{l}}\times L_{R_{l}}}$
instead of $\mathbf{I}$. The weighted sum power constraint is $\sum_{l=1}^{L}\textrm{Tr}\left(\mathbf{\Sigma}_{l}\hat{\mathbf{W}}_{l}\right)\leq P_{T}$,
where $\hat{\mathbf{W}}_{l}\in\mathbb{C}^{L_{T_{l}}\times L_{T_{l}}}$
is a positive definite weight matrix, $\forall l$. The whitened received
signal at receiver $R_{l}$ is a sufficient statistic,
\begin{eqnarray*}
\mathbf{y}_{l}' & = & \mathbf{W}_{l}^{-\frac{1}{2}}\mathbf{y}_{l}\\
 & = & \sum_{k=1}^{L}\left(\mathbf{W}_{l}^{-\frac{1}{2}}\mathbf{H}_{l,k}\hat{\mathbf{W}}_{l}^{-\frac{1}{2}}\right)\left(\hat{\mathbf{W}}_{l}^{\frac{1}{2}}\mathbf{x}_{k}\right)+\mathbf{W}_{l}^{-\frac{1}{2}}\mathbf{n}_{l}\\
 & = & \sum_{k=1}^{L}\mathbf{H}_{l,k}'\mathbf{x}_{k}'+\mathbf{n}_{l}',
\end{eqnarray*}
 which is the same as the received signal of a network with equivalent
channel state
\begin{eqnarray}
\mathbf{H}_{l,k}' & = & \mathbf{W}_{l}^{-\frac{1}{2}}\mathbf{H}_{l,k}\hat{\mathbf{W}}_{l}^{-\frac{1}{2}}\label{eq:equivalent_channel}
\end{eqnarray}
 and equivalent transmit signal $\mathbf{x}_{k}'=\hat{\mathbf{W}}_{l}^{\frac{1}{2}}\mathbf{x}_{k}$.
We name the left multiplication of $\mathbf{H}_{l,k}$ by $\mathbf{W}_{l}^{-\frac{1}{2}}$
as post-whitening and right multiplication of $\mathbf{H}_{l,k}$
by $\hat{\mathbf{W}}_{l}^{-\frac{1}{2}}$ as pre-whitening. Note that
$\mathbf{\Sigma}_{l}=\text{E}[\mathbf{x}_{l}\mathbf{x}_{l}^{\dagger}]$,
$\mathbf{\Sigma}_{l}'=\text{E}[\mathbf{x}_{l}'\left(\mathbf{x}_{l}'\right)^{\dagger}]=\hat{\mathbf{W}}_{l}^{\frac{1}{2}}\mathbf{\Sigma}_{l}\hat{\mathbf{W}}_{l}^{\frac{1}{2}}$,
and $\textrm{Tr}\left(\mathbf{\Sigma}_{l}\hat{\mathbf{W}}_{l}\right)=\textrm{Tr}\left(\hat{\mathbf{W}}_{l}^{\frac{1}{2}}\mathbf{\Sigma}_{l}\hat{\mathbf{W}}_{l}^{\frac{1}{2}}\right)=\textrm{Tr}\left(\mathbf{\Sigma}_{l}'\right)$.
Thus, the original network specified by
\begin{eqnarray*}
 &  & \left(\left[\mathbf{H}_{l,k}\right],\sum_{l=1}^{L}\textrm{Tr}\left(\mathbf{\Sigma}_{l}\hat{\mathbf{W}}_{l}\right)\leq P_{T},\left[\mathbf{W}_{l}\right]\right)
\end{eqnarray*}
 is equivalent to the network specified by
\begin{eqnarray}
 &  & \left(\left[\mathbf{W}_{l}^{-1/2}\mathbf{H}_{l,k}\hat{\mathbf{W}}_{k}^{-1/2}\right],\sum_{l=1}^{L}\textrm{Tr}\left(\mathbf{\Sigma}_{l}'\right)\leq P_{T},\left[\mathbf{I}\right]\right),\label{eq:equivalent-network}
\end{eqnarray}
 which has white noise and sum power constraint. 

Therefore, we can apply the dual link algorithm to the equivalent
network (\ref{eq:equivalent-network}) and find optimal $\mathbf{\Sigma}_{l}'$.
Then the solution to the original network is recovered by $\mathbf{\Sigma}_{l}=\hat{\mathbf{W}}_{l}^{-\frac{1}{2}}\mathbf{\Sigma}_{l}'\hat{\mathbf{W}}_{l}^{-\frac{1}{2}}$
because $\hat{\mathbf{W}}_{l}$ is non-singular.

\subsection{Distributed Implementation\label{sec:Distributed-Implementation}}

In practical applications, distributed algorithm with low coordination
overhead is highly desirable. It turns out that the dual link algorithm
is ideally suited for distributed implementation. 

A centralized implementation of the dual link incurs overhead. The
algorithm needs global channel state information $\mathbf{H}_{k,l},\ \forall k,l$,
as do other algorithms. The collection of global channel state information
wastes bandwidth and incur large delays for large networks. If the
delay is too long, there won't be enough time left for actual data
transmission before the channels change. 

Fortunately, a distributed implementation of the dual link algorithm
needs minimal local channel state information compared to other algorithms,
especially in TDD networks. A node only needs to estimate total received
signal covariance and desired link channel state information. No channel
state information of interfering links is needed. And the nodes do
not need to exchange channel state information because non-local CSI
is not needed. In a TDD network, assuming channel reciprocity, the
virtual reverse link iteration, Step 8 of Algorithm \ref{alg:1},
can be carried out in physical reverse links. Assume reverse link
$l$ transmits pilot signals using beamforming matrix $\hat{\mathbf{V}}_{l}$,
where $\hat{\mathbf{V}}_{l}\hat{\mathbf{V}}_{l}^{\dagger}=\hat{\boldsymbol{\Sigma}}_{l}$.
To calculate $\mathbf{\Sigma}_{l}$, we need $\hat{\mathbf{\Omega}}_{l}$
and reverse link total received signal covariance $\hat{\mathbf{\Omega}}_{l}+\mathbf{H}_{l,l}^{\dagger}\hat{\mathbf{\Sigma}}_{l}\mathbf{H}_{l,l}$,
which can be estimated at node $T_{l}$ because the channel has done
the summation of signal, interference, and noise for us for free,
no matter how large the network is. Using the pilot signal, $\mathbf{H}_{l,l}^{\dagger}\hat{\mathbf{V}}_{l}$
can be estimated and $\mathbf{H}_{l,l}^{\dagger}\hat{\mathbf{\Sigma}}_{l}\mathbf{H}_{l,l}=\mathbf{H}_{l,l}^{\dagger}\hat{\mathbf{V}}_{l}\hat{\mathbf{V}}_{l}^{\dagger}\mathbf{H}_{l,l}$
can be calculated and subtracted from $\hat{\mathbf{\Omega}}_{l}+\mathbf{H}_{l,l}^{\dagger}\hat{\mathbf{\Sigma}}_{l}\mathbf{H}_{l,l}$
to obtain $\hat{\mathbf{\Omega}}_{l}$. All reverse links only need
to share the scalar $\hat{\mu}$ in (\ref{eq:PWFS_reverse}) to adjust
the total power. The forward link calculation can be done similarly.
By avoiding global CSI collection, the distributed dual link algorithm
has significant lower signaling overhead compared to other methods.
The details of the distributed and scalable implementation and accommodation
of time varying channels will be presented in an upcoming paper \cite{Liu_2016_DistributedAlgorithmsWeightedSumrateMaximizationMIMOInterferenceNetworks}.

For distributed implementation in the case of colored noise and weighted
sum power constraint discussed in Section \ref{sec:Colored-Noise},
to create forward link $\mathbf{H}_{l,k}'$ in (\ref{eq:equivalent_channel}),
we can left multiply the transmit signal by $\hat{\mathbf{W}}_{k}^{-1/2}$
before transmission and right multiply the received signal by $\mathbf{W}_{l}^{-1/2}$.
In reverse links, it can be achieved by left multiplying the transmit
signal by $\mathbf{W}_{l}^{-1/2}$ before transmission and right multiplying
the received signal by $\hat{\mathbf{W}}_{k}^{-1/2}$. The dual link
algorithm assumes the reverse link channel noises are white. If the
noise covariance matrix of the reverse link $k$ of the TDD system
is $\hat{\mathbf{N}}_{k}$, we can estimate $\hat{\mathbf{W}}_{k}^{-1/2}\hat{\mathbf{N}}_{k}\hat{\mathbf{W}}_{k}^{-1/2}$
and replace it by $\mathbf{I}$ in the distributed algorithm.

\section{Simulation Results\label{sec:Simulation-Results}}

In this section, we provide numerical examples to compare the proposed
dual link algorithm with the PWF algorithm \cite{Liu_IT10s_Duality_BMAC},
the WMMSE algorithm \cite{Luo_TSP11_WMMSE}, and the iterative water-filling
algorithms for sum capacity of MAC and BC channels \cite{Yu_IT04_MIMO_MAC_waterfilling_alg,Jindal_IT05_IFBC}.
We consider a B-MAC network with $10$ data links among 10 transmitter-receiver
pairs that fully interfere with each other. Each link has 3 transmit
antennas and 4 receive antennas. For each simulation, the channel
matrices are independently generated and fixed by one realization
of $\mathbf{H}_{l,k}=\sqrt{g_{l,k}}\mathbf{H}_{l,k}^{\text{(W)}},\forall k,l$,
where $\mathbf{H}_{l,k}^{\text{(W)}}$ has zero-mean i.i.d. circularly
symmetric complex Gaussian entries with unit variance and $g_{l,k}$
is the average channel gain. The rate weights $w_{l}$'s are uniformly
chosen between 0.5 to 1. The total transmit power $P_{\text{T}}=100$.

\begin{figure}
\begin{centering}
\includegraphics[width=100mm]{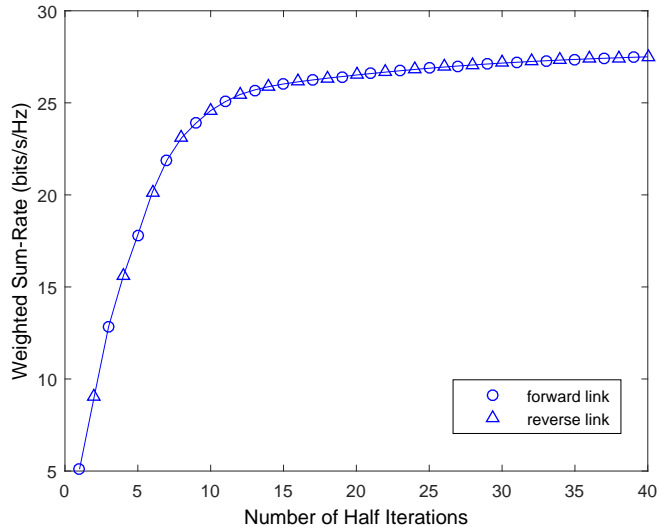}
\par\end{centering}

\caption{\label{fig:Convergence}The monotonic convergence of the forward and
reverse link weighted sum-rates of the Dual Link algorithm with $P_{\text{T}}=100$
and $g_{l,k}=0\text{dB},\forall l,k$.}
\end{figure}

Figure \ref{fig:Convergence} shows the convergence of the Dual Link
algorithm for a network with $g_{l,k}=0\text{dB},\forall l,k$. From
the proof of Theorem \ref{thm:monotone}, the weighted sum-rate of
the forward links and that of the reverse links not only increase
monotonically over iterations, but also increase over each other over
half iterations. In Algorithm \ref{alg:1}, the reverse link transmit
signal covariance matrices $\hat{\mathbf{\Sigma}}_{l}$ are updated
in the first half of each iteration, and the forward link transmit
signal covariance matrices $\mathbf{\Sigma}_{l}$ are updated in the
second half. From Figure \ref{fig:Convergence}, we see that the weighted
sum-rates of the forward links and reverse links increase in turns
until they converge to the same value, which also confirms that problem
(\ref{eq:WSRM_TP}) and its dual problem (\ref{eq:WSRM_DTP}) reach
their stationary points at the same time.

\begin{figure}
\begin{centering}
\textsf{\includegraphics[clip,width=100mm]{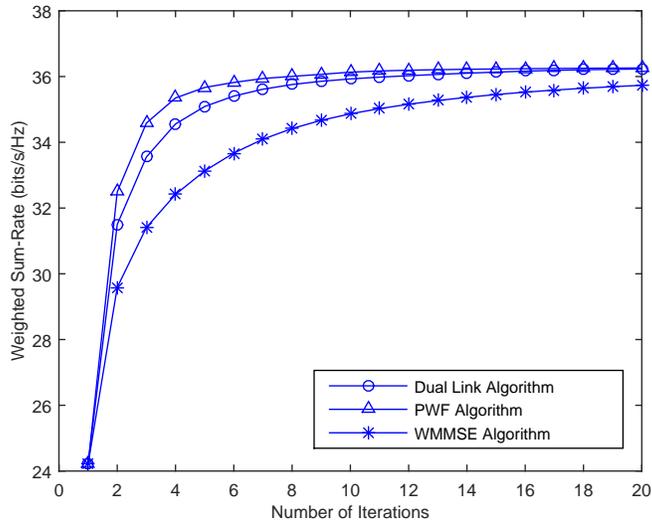}}
\par\end{centering}

\caption{\label{fig:weak_interference}PWF algorithm vs. WMMSE algorithm vs.
dual link algorithm under weak interference with $P_{\text{T}}=100$,
$g_{l,l}=0\text{dB}$ and $g_{l,k}=-10\text{dB}$ for $l\protect\ne k$.}
\end{figure}

\begin{figure}
\begin{centering}
\textsf{\includegraphics[clip,width=100mm]{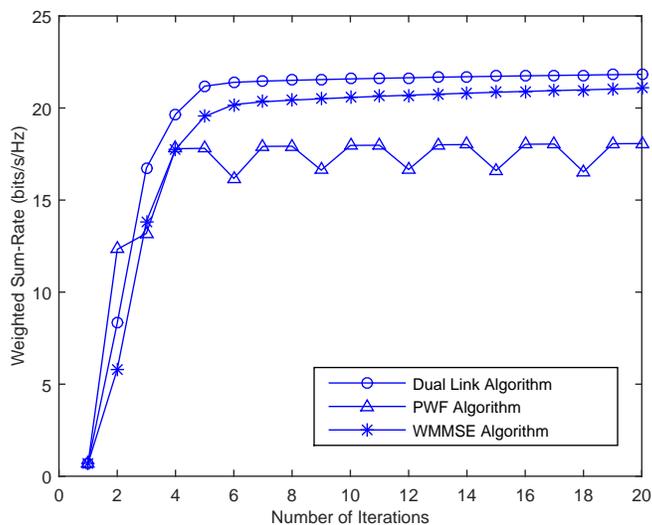}}
\par\end{centering}

\caption{\label{fig:strong_interference}PWF algorithm vs. WMMSE algorithm
vs. dual link algorithm under strong interference with $P_{\text{T}}=100$,
$g_{l,l}=0\text{dB}$ and $g_{l,k}=10\text{dB}$ for $l\protect\ne k$.}
\end{figure}

Figure \ref{fig:weak_interference} shows a typical case of rate versus
number of iterations under the weak interference setting ($g_{l,l}=0\text{dB}$
and $g_{l,k}=-10\text{dB}$ for $l\ne k$). For a channel realization
and a realization of random initial point, PWF algorithm converges
slightly faster than the dual link algorithm. In contrast, the WMMSE
algorithm's convergence is slower than the dual link algorithm, e.g.,
eight iterations to achieve what dual link algorithm achieves in four
iterations and more than ten iterations slower to reach some higher
value of the weighted sum-rate. When the gain of the interfering channels
are comparable to that of the desired channel, the difference in the
convergence speed between the PWF/Dual Link algorithm and the WMMSE
algorithm is less than that of the weak interference case but can
still be around five iterations for some high value of the weighted
sum-rate. Under strong interference setting, the PWF algorithm may
oscillate and no longer converge as shown in Figure \ref{fig:strong_interference}.
The dual link algorithm still converges slightly faster than the WMMSE
algorithm.

\begin{table}
\begin{centering}
\caption{\label{tab:avg}Average number of iterations needed to reach 90\%
and 95\% of the local optimum for PWF, WMMSE, and dual link algorithms.
$P_{\text{T}}=100$, $g_{l,l}=0\text{dB}$ and $g_{l,k}=-10/0/10\text{dB}$
for $l\protect\ne k$.}

\par\end{centering}

\centering{}%
\begin{tabular}{|c|c|c|c|}
\hline 
$g_{l,k}$ / Threshold & PWF & Dual Link & WMMSE\tabularnewline
\hline 
\hline 
-10 dB / 90\% & 1.199 & 1.653 & 2.521\tabularnewline
\hline 
0 dB / 90\% & 6.075 & 7.211 & 9.557\tabularnewline
\hline 
10 dB / 90\% & 5.217 & 4.745 & 5.847\tabularnewline
\hline 
-10 dB / 95\% & 2.140 & 2.781 & 5.067\tabularnewline
\hline 
0 dB / 95\% & 11.465 & 12.408 & 15.864\tabularnewline
\hline 
10 dB / 95\% & 9.358 & 6.837 & 10.549\tabularnewline
\hline 
\end{tabular}
\end{table}

Table \ref{tab:avg} shows the average convergence speed of these
three algorithms under different interference levels. The results
are obtained by averaging over 1000 independent channel realizations
under each interference setting. Under the strong interference setting
($g_{l,k}=10\text{dB}$), only the converged cases (837 out of 1000)
are considered for the PWF algorithm. We can see that the dual link
algorithm outperforms WMMSE algorithm in all the settings. While the
PWF algorithm is slightly ahead under weak and moderate interference
settings, it is slower than the dual link algorithm under the strong
interference setting. 

Given the same initial point, these three algorithms may converge
to different stationary points. Since the original weighted sum-rate
maximization problem is non-convex, a stationary point is not necessarily
a global maximum. In practical applications, one may run such algorithms
multiple times starting from different initial points and choose the
best.

\begin{figure}
\begin{centering}
\includegraphics[width=100mm]{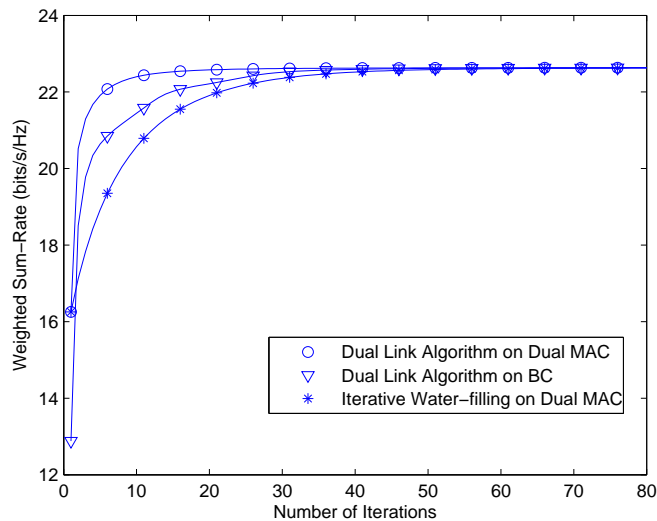}
\par\end{centering}

\caption{\label{fig:MAC_BC}Dual Link algorithm vs. Iterative water-filling
on MAC and BC channels}
\end{figure}

Figure $\ref{fig:MAC_BC}$ compares the dual link algorithm with the
iterative water-filling algorithm proposed in \cite{Yu_IT04_MIMO_MAC_waterfilling_alg,Jindal_IT05_IFBC}
for sum capacity of BC and MAC channels. The iterative water-filling
algorithm of \cite{Yu_IT04_MIMO_MAC_waterfilling_alg,Jindal_IT05_IFBC}
works because that the sum rate can be written in a single user rate
form and because of the duality between MAC and BC channels. We consider
a BC channel and its dual MAC channel with a sum power constraint.
The BC channel contains 1 transmitter and 10 receivers, each equipped
with 5 antennas. Entries of the channel matrices are generated from
i.i.d. zero mean Gaussian distribution with unit variance. The noise
covariance is an identity matrix. As shown in Figure $\ref{fig:MAC_BC}$,
the dual link algorithm and the iterative water-filling algorithm
all converge to the same sum-capacity. The dual link algorithm converges
significantly faster than the iterative water-filling algorithm.

\section{Complexity Analysis\label{sec:Complexity-Analysis}}

We have numerically evaluated the convergence properties of the proposed
new algorithm, the PWF algorithm and the WMMSE algorithm in terms
of the number of iterations. We now analyze the complexity per iteration
for these algorithms. 

Recall that $L$ is the number of users or links, and for simplicity,
assume that each user has $N$ transmit and $N$ receive antennas,
so the resulting $\mathbf{\Sigma}_{l}$ (and $\hat{\mathbf{\Sigma}}_{l}$)
is a $N\times N$ matrix. Suppose that we use the straightforward
matrix multiplication and inversion. Then the complexity of these
operations are $O(N^{3})$. For the dual link algorithm, at each iteration,
the calculation of $\mathbf{\Omega}_{l}$ incurs a complexity of $O(LN^{3})$
and the calculation of $\mathbf{\Omega}_{l}+\mathbf{H}_{l,l}\mathbf{\Sigma}_{l}^{(n+1)}\mathbf{H}_{l,l}^{\dagger}$
incurs a complexity of $O(LN^{3})$. To obtain $\hat{\mathbf{\Sigma}}_{l}$,
we have to invert two $N\times N$ matrices, which incurs a complexity
of $O(N^{3})$. Therefore, the total complexity for calculating a
$\hat{\mathbf{\Sigma}}_{l}$ is $O(LN^{3})$, and the complexity of
generating all $\hat{\mathbf{\Sigma}}$ is $O(L^{2}N^{3})$. As calculating
$\mathbf{\Sigma}$ incurs the same complexity as calculating $\hat{\mathbf{\Sigma}}$,
the complexity of the new algorithm is $O(L^{2}N^{3})$ for each iteration. 

The PWF algorithm uses the same calculation to generate $\mathbf{\Omega}_{l}$
and incurs a complexity of $O(LN^{3})$ for each $\mathbf{\Omega}_{l}$.
Then, it uses the singular value decomposition of $\mathbf{\Omega}_{l}^{-\frac{1}{2}}\mathbf{H}_{l,l}\hat{\mathbf{\Omega}}_{l}^{-\frac{1}{2}}$,
which incurs a complexity of $O(N^{3})$. Since we need $L$ of these
operations, the total complexity of the PWF algorithm is $O(L^{2}N^{3})$.
For the WMMSE algorithm, it is shown in \cite{Luo_TSP11_WMMSE} that
its complexity is $O(L^{2}N^{3})$. 

So, all three algorithms have the same computational complexity per
iteration if we use $O(N^{3})$ matrix multiplication. Recently, Williams
\cite{williams2012multiplying} presents an $O(N^{2.3727})$ matrix
multiplication and inversion method. If we use this algorithm, then
the new algorithm and the WMMSE algorithm have $O(L^{2}N^{2.3727})$
complexity since the $N^{3}$ factor comes from the matrix multiplication
and inversion. However, in addition to $L^{2}$ number of matrix multiplications
and inversions, the PWF algorithm has $L$ number of $N$ by $N$
matrix singular value decompositions. Therefore the complexity of
the PWF algorithm is $O(L^{2}N^{2.3727}+LN^{3})$.

\section{Conclusion\label{sec:Conclusion}}

We have proposed a new algorithm, the dual link algorithm, to solve
the weighted sum-rate maximization problem in general interference
channels. Based on the polite water-filling results and the rate duality
\cite{Liu_IT10s_Duality_BMAC}, the proposed dual link algorithm updates
the transmit signal covariance matrices in the forward and reverse
links in a symmetric manner and has fast and guaranteed convergence.
We have given a proof for its monotonic convergence, and the proof
technique may be generalized for other problems. Numerical examples
have demonstrated that the new algorithm has a convergence speed close
to the fastest PWF algorithm which is however not guaranteed to converge
in all situations. The dual link algorithm is scalable and well suited
for distributed implementation. It can also be easily modified to
accommodate colored noise.

\bibliographystyle{plain}
\bibliography{liu_research,xingl}

\begin{thebibliography}{10}

\bibitem{Jafar_2008ITIT_InterferenceAlignmentDegreesFreedomUserInterferenceChannel}
V.R. Cadambe and S.A. Jafar.
\newblock Interference {{Alignment}} and {{Degrees}} of {{Freedom}} of the
  {{K-User Interference Channel}}.
\newblock {\em IEEE Transactions on Information Theory}, 54(8):3425--3441,
  August 2008.

\bibitem{ChenL_CoRR2013_WSR_MIMO_minimax}
L.~Chen and S.~You.
\newblock The weighted sum rate maximization in {MIMO} interference networks:
  The minimax lagrangian duality and algorithm.
\newblock {\em IEEE Trans. Networking}, submitted, Sep., 2013.

\bibitem{Caire_09ISIT_BC_linear_constraints}
H.~Huh, H.~Papadopoulos, and G.~Caire.
\newblock {MIMO} broadcast channel optimization under general linear
  constraints.
\newblock In {\em Proc. IEEE Int. Symp. on Info. Theory (ISIT)}, 2009.

\bibitem{Caire_09sTSP_BC_intercell_interference}
H.~Huh, H.~C. Papadopoulos, and G.~Caire.
\newblock Multiuser {MISO} transmitter optimization for intercell interference
  mitigation.
\newblock {\em IEEE Transactions on Signal Processing}, 58(8):4272 --4285,
  August 2010.

\bibitem{Jindal_IT05_IFBC}
N.~Jindal, W.~Rhee, S.~Vishwanath, S.~A. Jafar, and A.~Goldsmith.
\newblock Sum power iterative water-filling for multi-antenna {Gaussian}
  broadcast channels.
\newblock {\em IEEE Trans. Info. Theory}, 51(4):1570--1580, April 2005.

\bibitem{Giannakis_TIT11_IFPMIMO}
Seung-Jun Kim and G.B. Giannakis.
\newblock Optimal resource allocation for {MIMO} ad hoc cognitive radio
  networks.
\newblock {\em IEEE Trans. Info. Theory}, 57(5):3117 -- 3131, May 2011.

\bibitem{Liu_2016_DistributedAlgorithmsWeightedSumrateMaximizationMIMOInterferenceNetworks}
Xing Li, Lijun Chen, Behrouz Touri, Xinming Huang, and Youjian~Eugene Liu.
\newblock Distributed {{Algorithms}} for {{Weighted Sum}}-rate {{Maximization}}
  in {{MIMO Interference Networks}}.
\newblock to be submitted, 2016.

\bibitem{Liu_10sTSP_Fairness_rate_polit_WF}
An~Liu, Youjian Liu, Haige Xiang, and Wu~Luo.
\newblock {MIMO B-MAC} interference network optimization under rate constraints
  by polite water-filling and duality.
\newblock {\em IEEE Trans. Signal Processing}, 59(1):263 --276, January 2011.

\bibitem{Liu_10sTSP_MLC}
An~Liu, Youjian Liu, Haige Xiang, and Wu~Luo.
\newblock Polite water-filling for weighted sum-rate maximization in {B-MAC}
  networks under multiple linear constraints.
\newblock {\em IEEE Trans. Signal Processing}, 60(2):834 --847, February 2012.

\bibitem{Liu_IT10s_Duality_BMAC}
An~Liu, Youjian Liu, Haige Xiang, and Wu~Luo.
\newblock Duality, polite water-filling, and optimization for {MIMO B-MAC}
  interference networks and {iTree} networks.
\newblock {\em IEEE Trans. Info. Theory}, in revision, submitted Apr. 2010.

\bibitem{Berry_MonoIFCpricing_ISIT09}
C.~Shi, R.~A. Berry, and M.~L. Honig.
\newblock Monotonic convergence of distributed interference pricing in wireless
  networks.
\newblock {\em in Proc. IEEE ISIT, Seoul, Korea}, June 2009.

\bibitem{Luo_TSP11_WMMSE}
Qingjiang Shi, M.~Razaviyayn, Zhi-Quan Luo, and Chen He.
\newblock An iteratively weighted mmse approach to distributed sum-utility
  maximization for a mimo interfering broadcast channel.
\newblock {\em IEEE Trans. Signal Processing}, 59(9):4331 --4340, September
  2011.

\bibitem{williams2012multiplying}
Virginia~Vassilevska Williams.
\newblock Multiplying matrices faster than coppersmith-winograd.
\newblock In {\em Proceedings of the 44th symposium on Theory of Computing},
  pages 887--898. {ACM}, 2012.

\bibitem{YouS_14sGlobecom_cvxccv_short}
Seungil You, Lijun Chen, and Y.E. Liu.
\newblock Convex-concave procedure for weighted sum-rate maximization in a
  {{MIMO}} interference network.
\newblock In {\em 2014 {{IEEE Global Communications Conference}}
  ({{GLOBECOM}})}, pages 4060--4065, December 2014.

\bibitem{Yu_IT06_Minimax_duality}
W.~Yu.
\newblock {Uplink-downlink duality via minimax duality}.
\newblock {\em IEEE Trans. Info. Theory}, 52(2):361--374, 2006.

\bibitem{Yu_IT04_MIMO_MAC_waterfilling_alg}
W.~Yu, W.~Rhee, S.~Boyd, and J.M. Cioffi.
\newblock Iterative water-filling for {Gaussian} vector multiple-access
  channels.
\newblock {\em IEEE Trans. Info. Theory}, 50(1):145--152, 2004.

\bibitem{Weiyu_IT06_DualIWF}
Wei Yu.
\newblock Sum-capacity computation for the gaussian vector broadcast channel
  via dual decomposition.
\newblock {\em IEEE Trans. Inform. Theory}, 52(2):754--759, February 2006.

\bibitem{Wei_07ITW_MultiuserWF}
Wei Yu.
\newblock Multiuser water-filling in the presence of crosstalk.
\newblock {\em Information Theory and Applications Workshop, San Diego, CA,
  U.S.A}, pages 414 --420, 29 2007-feb. 2.

\bibitem{Zhang_09ISIT_BC_MAC_duality_linear_constraints}
Lan Zhang, Ying-Chang Liang, Yan Xin, Rui Zhang, and H.V. Poor.
\newblock On gaussian {MIMO BC-MAC} duality with multiple transmit covariance
  constraints.
\newblock In {\em Proc. IEEE Int. Symp. on Info. Theory (ISIT)}, pages 2502
  --2506, June 2009.

\bibitem{ZhangLan_09TWC_Weighted_rate_BC}
Lan Zhang, Yan Xin, and Ying-chang Liang.
\newblock Weighted sum rate optimization for cognitive radio {MIMO} broadcast
  channels.
\newblock {\em IEEE Trans. Wireless Commun.}, 8(6):2950 --2959, June 2009.

\end{thebibliography}

\end{document}